# Resonant splitting in periodic T-shaped photonic waveguides

**Wenjin Yin,[1] Kaike Yang,[1] Yuee Xie,[1] Tao Ouyang,[1] Leo Lau[2] and Yuanping Chen*[1]**

[1]*Department of Physics and Laboratory for Quantum Engineering and Micro-Nano Energy Technology, Xiangtan University, Xiangtan 411105, Hunan, China*

[2]*Beijing Computational Science Research Center, Beijing 100084, China*

*Corresponding author: chenyp@xtu.edu.cn*

**Abstract:** Resonant splitting phenomena of photons in periodic T-shaped waveguide structure are investigated by the Green's function method. An interesting resonant phenomenon is found in the transmission spectra. When the T-shaped structure contains $n$ constrictions, there are ($n$-1)-fold resonant splitting peaks in the low frequency region. The peaks are induced by low quasi-bound states in which photons are intensively localized in the stubs. While ($n$-2)-fold resonant splitting occurs in the high frequency region. These peaks are induced by the high quasi-bound states where the photons are mainly localized in constrictions rather than in stubs. To this kind of quasi-bound state, the stub acts as a potential barrier rather than a well, which is the inverse of the case of the low quasi-bound states corresponding to the ($n$-1)-fold splitting peaks. These resonant peaks can be modulated by adjusting the periodic number and geometry of the waveguide structures. The results can be used to develop optical switching devices, tunable filters, and coupled waveguides.

# I    INTRODUCTION

With the development of nanofabrication and lithography technology, various kinds of low dimensional periodic optical structures have been fabricated.[1-5] Photonic waveguides, as typical optical structure, have received recent much theoretical[6-8] and experimental[9-11] attentions, due to their ability to localize light and control its propagation similar to the semiconductor material's control of electrons. So far, many impressive photonic waveguides have been explored and have shown exciting properties. Hetero-structure photonic waveguides, which are formed by embedding two or more dielectric films or photonic crystals into a single structure, has been studied by both theoretical and experimental.[12-14] It has been found that the number of bound states within the waveguides depended on the width and well depth and the Hetero-structure have been used to develop devices such as resonant cavities[15], waveguides[16] and filters[17]. Meanwhile nonlinear nanophotonic coupling waveguides induced by laser or stress field have also been investigated both theoretically[18-20] and experimentally[21-22]. Nonlinear photonic waveguides are formed by embedding Kerr-nonlinear photonic crystals in a linear photonic crystal. The bound states split into symmetric and antisymmetric pairs due to the coupling between waveguides. In addition, for a multiple photonic well waveguides consisting of $n$ photonic potential barriers and $n$-1 potential wells, ($n$-1)-fold resonant splitting appears in the transmission spectra.[23-26] These peaks are induced by the bound states strongly localized in the potential wells. The total number of the bound states can be controlled by varying the periodicity and the width of the wells, and the bound states have been used to develop multiple channeled filters and other optical devices.[9-11] However, all the results mentioned above were corresponding to the photonic waveguides with potential barriers or wells. Recently, the photonic resonant transmission phenomenon is also found in open periodic waveguide structure where no potential barriers exist in the quantum channel.[27] A periodic waveguide structure can be formed by cascading uniform waveguide parts with alternating widths where the narrow section is called a constriction while the wide region between two adjacent constrictions is a stub.[28] As photons transport across the periodic waveguide structure, it has been shown that ($n$-1)-fold resonant splitting occurs in the photonic band gap.[27] The peaks are caused by the quasi-bound states where the photons are mainly confined in the stubs. Compared with the waveguides structure mentioned above, the stub in the open periodic waveguide structure is equivalent to a potential well, while the constriction is regarded as a barrier. However, the constrictions in open periodic structure are not real potential barriers after all. Sometimes carriers in the open structure can be confined in the constrictions.[29] Thus, the open periodic waveguide structure should possess some exotic transport properties.

In this paper, we study photonic transport in a typical open periodic waveguide structure periodic T-shaped waveguide, as shown in Fig. 1. It is interesting to find that in the transmission spectra not only there are ($n$-1)-fold resonant splitting peaks in the low frequency region but also there exist ($n$-2)-fold resonant splitting peaks in the high frequency region. Moreover, ($n$-2)-fold splitting peaks are induced by the high quasi-bound states strongly confined in the constrictions rather than in the stubs, which is the inverse of the case of the low quasi-bound states corresponding to the ($n$-1)-fold splitting peaks. To the photons confined in the high quasi-bound states, the stub acts as a potential barrier rather than a well.

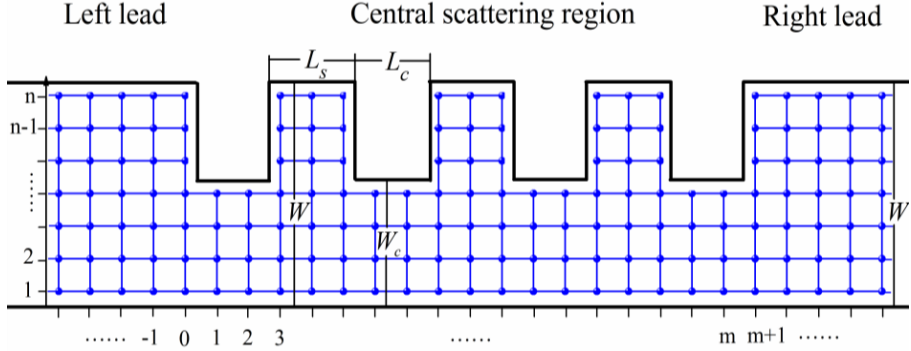

Fig. 1. Schematic view of the periodic T-shaped photonic waveguide structure, where the system is divided into a set of effective square lattices with lattice constant $\nabla$. It consists of a finite periodic structure connected to two semi-infinite leads with width $W$. An unit cell of the finite periodic structure consists of a stub with size $W \times L_s$ and a constriction with size $W_c \times L_c$.

## II  MODEL AND METHOD

The periodic T-shaped waveguide we considered is as depicted in Fig. 1. It consists of two semi-infinite leads (Left lead and Right lead) with the width $W$ connected by the central sandwiched scattering region in horizontal direction. The central region is a finite periodic structure which consists of constrictions with size $W_c \times L_c$ and stubs with size $W \times L_s$. The lattice Green's function technique[30,31] can be used to calculate the photonic transport properties of the waveguide structure. In terms of the Green's function scheme, the system is divided into a set of effective square lattices with lattice constant $\nabla$. To describe the photonic properties of the effective discretized system, the tight-binding approximation Hermitian operators are used and expressed as:[30]

$$\hat{H} = \sum_{m,n}(\upsilon_{m,n}a^+_{m,n}a_{m,n} - u_{m,m+1;n,n}a^+_{m,n}a_{m+1,n} - u_{m+1,m;n,n}a^+_{m+1,n}a_{m,n}$$
$$- u_{m,m;n,n+1}a^+_{m,n}a_{m,n+1} - u_{m,m;n+1,n}a^+_{m,n+1}a_{m,n}), \quad (1)$$

where $a^+_{m,n}$, $a_{m,n}$ are creation and annihilation operators at $(m, n)$ site, the $\upsilon_{m,n}$ represents the site energy, $u_{m,m+1;n,n}$ and $u_{m+1,m;n,n}$ denote forward and backward hopping energy between two neighboring sites in the horizontal direction, and $u_{m,m;n,n+1}$ and $u_{m,m;n+1,n}$ describe similar hopping energy in the vertical direction. In the case of transverse magnetic mode, the coefficients can be defined as follows:

$$\upsilon_{m,n} = 4\xi^2_{m,n}, \quad u_{m,m+1;n,n} = \xi_{m,n}\xi_{m+1,n}, \quad u_{m+1,m;n,n} = \xi_{m+1,n}\xi_{m,n},$$
$$u_{m,m;n+1,n} = \xi_{m,n+1}\xi_{m,n}, \quad u_{m,m;n,n+1} = \xi_{m,n}\xi_{m,n+1}, \quad \xi_{m,n} = \varepsilon^{-1/2}_{m,n}, \quad (2)$$

here, $\varepsilon_{m,n}$ is the system's dielectric constant.

Based on the Hermitian operator $\hat{H}$ in Eq. (1), the Green's function of the whole system can be expressed as:[30]

$$G^r(\omega) = \left((\omega/c)^2 - H_S - \Sigma_L - \Sigma_R\right)^{-1}, \quad (3)$$

where $c$ is the velocity of light, $\omega$ is the photon frequency, and $H_S$ is the Hemitian operator of central scatting region. $\Sigma_{L(R)} = \hat{h}_{SL(R)}g^r_{L(R)}\hat{h}_{L(R)S}$ denotes the self-energy for leads. $\hat{h}_{SL} = [h_{LS}]^+$ and $\hat{h}_{SR} = [h_{RS}]^+$ are the coupling matrix, which can be written as:

$$h_{SL} = -\sum_n\left(u_{1,0;n,n}a^+_{1,n}a_{0,n}\right), \quad (4)$$

$$h_{SR} = -\sum_n(u_{m,m+1;n,n}a^+_{m,n}a_{m+1,n}), \quad (5)$$

$g^r_{L(R)}$ is the surface Green's function which can be computed by recursive iteration technique[30,31]. Once the Green's function is obtained, one can calculate the transmission coefficient and the photonic *LDOS*

$$T(\omega) = Tr\left[G^r(\omega)\Gamma_L G^a(\omega)\Gamma_R\right], \tag{6}$$

$$LDOS(\omega) = -\text{Im}\left[G^r(\omega)\right]/\pi, \tag{7}$$

where $G^a(\omega) = \left[G^r(\omega)\right]^+$, and $\Gamma_{L(R)} = i\left(\Sigma^r_{L(R)} - \Sigma^a_{L(R)}\right)$ describes the interaction between the left or right lead and the central scattering region.

### III  RESULTS AND DISCUSSION

In Fig. 2, the photonic transmissions of the periodic T-shaped waveguide structures with different periodic number *n* are shown. One can find that there are some sharp resonant splitting peaks in the spectra which can be divided into two categories according to their locations. The splitting peaks of one category locate in the low frequency region (0.045~0.047), while the others lie in the high frequency region slightly higher than the first threshold frequency (the omega is about 0.059). The two categories of resonant splitting peaks present different splitting rules. For the waveguide consisting of two constrictions and a single stub, *i.e.*, *n*=2, there only exists one resonant peak in the low frequency region (see Fig. 2(a)). In the case of the waveguide consisting of three constrictions and two stubs, *i.e.*, *n*=3, the peak splits into two in the low frequency region, meanwhile a sharp peak appears in the high frequency as shown in Fig. 2(b). As the periodic number of the waveguide structures increases to *n*=4 and 5, more and more resonant splitting peaks emerge both in low and high frequency regions (see Figs. 2(c) and 2(d)). By checking the number of the peaks in different frequency region, it is clear to find that the first category of resonant peaks in the low frequency region satisfies (*n*-1)-fold splitting rule, while the second category in the high frequency region satisfies (*n*-2)-fold splitting rule.

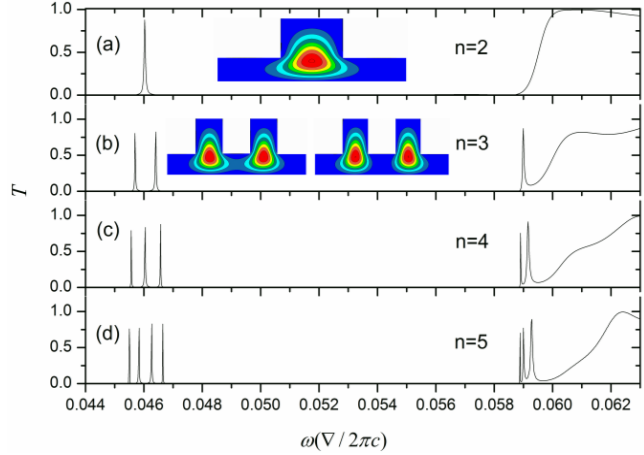

Fig. 2. (Color online) Transmission coefficient ($T$) versus photon frequency ($\omega$) for the periodic T-shaped waveguide structure in Fig. 1 with different periodic number $n$. (a) $n=2$, (b) $n=3$, (c) $n=4$ and (d) $n=5$. The structure parameters are chosen to be $W=20\nabla$, $L_s=10\nabla$, $W_c=8\nabla$, $L_c=10\nabla$. Inset: (a) the photonic $LDOS$ low quasi-bound states corresponding to the resonant peak, (b) left and right insets are photonic $LDOS$ of the low quasi-bound states correspond to the first and second peaks.

As mentioned above, the ($n$-1)-fold resonant splitting peaks in low frequency region are caused by the quasi-bound states where the photons are mainly confined in the stubs. This point can be easily confirmed by the photonic $LDOS$ of the quantum states corresponding to these peaks, as shown in the insets of Figs. 2(a) and 2(b). To these low quasi-bound states, each stub is equivalent to a potential well, while each constriction is a potential barrier. As the waveguide consists of $n$-1 stubs and $n$ constrictions, there exist $n$-1 potential wells and $n$ potential barriers in the structure. The coupling effect among the $n$-1 potential wells leads to $n$-1 low quasi-bound states. Therefore, ($n$-1)-fold resonant splitting peaks are exhibited in the low frequency region of transmission spectra.

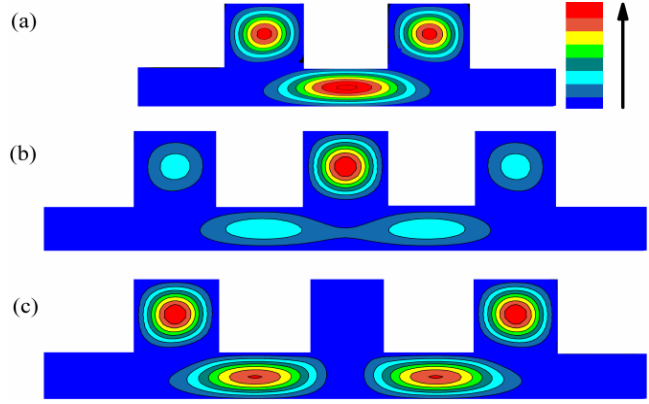

Fig. 3. (Color online) The photonic *LDOS* of high quasi-bound states corresponding to the resonant splitting peaks in the high frequency region in Fig. 2. (a) Quasi-bound state corresponding to the third resonant peak in Fig. 2(b); (b), (c) Quasi-bound states corresponding to the fourth and the fifth resonant peaks in Fig. 2(c), respectively.

To explain the origination of the (*n*-2)-fold resonant splitting in Fig. 3, we present the photonic *LDOS* of quantum states corresponding to the resonant peaks in the high frequency region. Figure 3(a) shows the photonic *LDOS* of quantum state corresponding to the third peak in Fig. 2(b). One can see that in the longitudinal direction the photons are mainly localized in the middle constriction rather than the stub, which is different from the low quasi-bound state in the inset of Fig. 2(a). Figure 3(b) and 3(c) are the photonic *LDOS* of quantum states corresponding to the fourth and fifth splitting peaks in Fig. 2(c), respectively. It can be found that the photons in the two states are still confined in the middle constrictions. Moreover, the two high quasi-bound states are the result of the symmetric and anti-symmetric super-positions of two single high quasi-bound states in Fig. 3(a). To these high quasi-bound states, the stubs are equivalent to potential barriers rather than wells, while the middle constrictions are equivalent to wells. Obviously, this is the inverse of the case of the low quasi-bound state. Therefore, when the periodic T-shaped waveguide consists of $n$-1 stubs and $n$ constrictions, there exist $n$-1 potential barriers and $n$-2 wells in the structure. The interaction among the quantum states in the $n$-2 potential wells leads to $n$-2 high quasi-bound states, and thus ($n$-2)-fold resonant splitting peaks appear in the high frequency region. It should be mentioned that, for a waveguide consists of only two constrictions, *i.e.*, $n$=2, there is no peak in the high frequency region (see in Fig. 2(a)). This is because the photons in the two constrictions will escape to the left or right lead and thus no quasi-bound states can be formed in the structure.

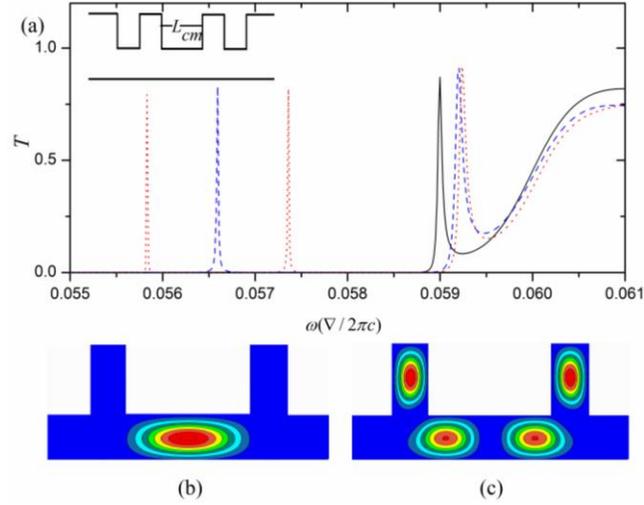

Fig. 4. (Color online) (a) Transmission coefficient (*T*) versus photon frequency ($\omega$) for the structure in the inset with different length $L_{cm}$ of the middle constriction. The solid, dashed and dotted lines represent the transmission curves at $L_{cm}$=10$\nabla$, 32$\nabla$, and 55$\nabla$, respectively. Inset: Similar structure as in Fig. 1 with three constrictions and two stubs; $L_{cm}$ is the length of the middle constriction. (b), (c) the photonic *LDOS* of high quasi-bound states corresponding to the first and the second resonant peaks of the dashed line in (a), respectively. Other parameters are the same as in Fig. 2.

To further explain the effect of the constriction on the high quasi-bound states, we study the resonant transmission for the waveguide consisting of three constrictions and two stubs as shown in the inset of Fig. 4(a). Figure 4(a) presents the transmission spectra for the structure with different length $L_{cm}$ of the middle constriction. One can find that the resonant splitting peaks induced by high quasi-bound states gradually shift to the low frequency region as the length $L_{cm}$ increases, meanwhile the number of resonant peaks increases, *i.e.*, the number of high quasi-bound states increases with the length $L_{cm}$. Figures 4(b) and 4(c) present the photonic *LODS* of the two high quasi-bound states corresponding to the first and the second resonant splitting peaks of the dashed line in Fig. 4(a), respectively. In the longitudinal direction, these high quasi-bound states are mainly localized in the middle constriction. Furthermore, one can find that the two states are respectively even and odd symmetric with respect to the center line of the middle constriction, which is similar to the case of a quasi-one-dimensional double barriers structure. This further identifies that, for the high quasi-bound states, the middle constriction sandwiched in the two stubs indeed acts as a potential well rather than a potential barrier.

## IV CONCLUSION

In summary, we have investigated the photon transport properties of periodic T-shaped waveguide structure by using the Green's function method. It has been found that ($n$-1)-fold resonant splitting peaks appear in the low frequency region, while ($n$-2)-fold splitting occurs in high frequency region of the transmission spectra. The ($n$-1)-fold resonant splitting peaks are induced by low quasi-bound states strongly localized in the stubs. However, the (*n*-2)-fold splitting peaks are originated from the high quasi-bound states mainly localized in the constrictions. To the photons confined in the high quasi-bound states, the constrictions act as potential wells rather than barriers. More quasi-bound states will exist in the constriction as the constriction length increases. The results can be used to develop optical switching devices, tunable filters, and coupled waveguides.

## ACKNOWLEDGMENTS

This work was supported by the National Natural Science Foundation of China (Nos. 51006086, 11074213, 51176161).